\documentclass[preprint,aps]{revtex4}
\usepackage{graphics}
\usepackage{graphicx}
\usepackage{dcolumn}
\usepackage{bm}

\begin{document}

\title{Break-junction experiments on the zero-bias anomaly of non-magnetic and
ferromagnetically ordered metals}

\author{Kurt Gloos$^{1,2}$ and Elina Tuuli$^{1-3}$}

\address{$^1$ Wihuri Physical Laboratory, Department of Physics and
Astronomy, University of Turku, FIN-20014 Turku, Finland}

\address{$^2$ Turku University Centre for Materials and Surfaces 
  (MatSurf), FIN-20014 Turku, Finland}

\address{$^3$ Graduate School of Materials Research (GSMR), FIN-20500 Turku, Finland}


\begin{abstract}
We have investigated break junctions of normal non-magnetic metals as 
well as ferromagnets at low temperatures.
The point contacts with radii $0.15 - 15\,$nm showed zero-bias anomalies
which can be attributed to Kondo scattering at a single Kondo impurity at the 
contact or to the switching of a single conducting channel. The Kondo 
temperatures derived from the width of the anomalies varied between 10 and 
1000\,K. 
These results agree well with literature data on atomic-size contacts of the 
ferromagnets as well as with spear-anvil type contacts on a wide variety of 
metals.
\end{abstract}

\maketitle 

\section{Introduction}
The recent discovery of the Kondo effect in atomic-size contacts of the ferromagnets 
Co, Fe, and Ni demonstrates how material properties can be changed by reducing the 
size of the sample. In that specific case magnetic order was quenched at the junction,
enabling the Kondo effect to take place \cite{Calvo2009}.
Kondo scattering also seems to be behind the so-called zero-bias anomalies
of point contacts of a wide variety of metals including magnets, nonmagnets, 
and even superconductors \cite{Gloos2009}.
The latter experiments revealed a systematic variation of the zero-bias anomalies
as function of contact size.
However, spear-anvil type contacts usually involve interfaces which 
can be degraded by oxidation or a residual H$_2$O film, which
is avoided by using mechanically controllable break junctions. 
Here we report break junction experiments on the ferromagnet Fe and compare
them to those of non-magnetic Cu.
Similar results were obtained for Co, Ni, Al, and Cd. 

The resistance $ R(T) \approx 2R_K/(ak_F)^2 + \rho(T)/(2a)$ of a point contact 
between two identical normal metals  depends on the Fermi wave number 
$k_F$, the contact radius $a$, and the temperature-dependent specific resistivity 
$\rho(T)$  \cite{Naidyuk2005}. Here $R_K = h/e^2$.
In the ballistic limit electrons cross the contact region on straight trajectories 
and the first term dominates.
The second term provides corrections due to electron scattering in the contact 
region, the so-called backscattering.
According to the Drude-Sommerfeld theory \cite{Ashcroft1976} the
product of electrical resistivity and electron mean free path $\rho \cdot l =
\left(3\pi R_K \right) / \left( 2 k_F^2 \right)$ is a constant. 
Therefore  the temperature dependence of the contact resistance can be replaced by 
the energy or bias-voltage dependence at low temperature 
\begin{equation}
  \frac{dV}{dI} \approx \frac{2R_K}{(ak_F)^2}\left(1 + \frac{a}{l(eV)} \right)
  \label{pc-backscatter}
  \end{equation}
to extract $l(eV)$. 
Independent scattering processes can be separated if their respective mean 
free paths have different energy dependencies. 
This is the case for electron-phonon scattering, which sets in at energies 
above $5 - 10\,$meV, and Kondo scattering at magnetic impurities that is 
efficient only at small  energies.
A Kondo impurity polarizes the surrounding electrons,  forming
a  $\xi_K \approx \hbar v_F / k_B T_K$ large polarization cloud. 
Here $v_F$ is the Fermi velocity.
The electrical resistivity \cite{Hamann1967}
\begin{equation}
  \Delta \rho(T) = \frac{\Delta \rho_0}{2} \left[
               1 - \frac{\ln{\left( T/T_K \right)}} {\sqrt{\ln^2{\left(\ T/T_K \right)} + S(S+1)\pi^2  } }
  \right]
  \label{pc-kondo}
  \end{equation}
due to scattering of electrons at Kondo impurities 
depends on the Kondo temperature $T_K$ and the effective spin $S$.
It reaches a maximum (in the unitary limit) of $\Delta \rho_0 = c \cdot 2 R_K/k_F$ at 
low temperatures
proportional to the impurity concentration $c$ (impurities per conduction electron).
Even a single magnetic impurity can change dramatically the  resistance of a 
point contact.

\section{Experimental and results}

\begin{figure}
  \includegraphics{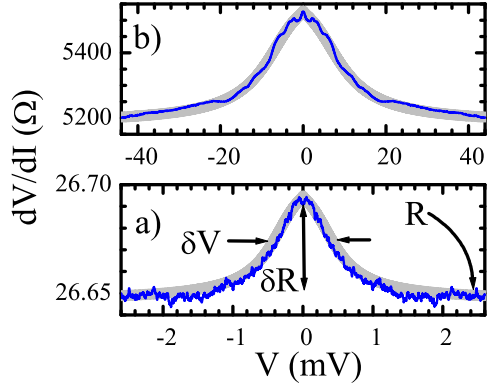}
  \caption{\label{kondo-fit}
  Differential resistance of two typical Fe contacts at 1\,K.
  Note the different scales. The spectra have been symmetrized.
  The thick grey lines are fits using Equation \ref{pc-kondo} with 
  $S=0.1$.
  The lower Figure shows definitions for the contact resistance $R$, the 
  magnitude $\delta R$, and   the width $\delta V$ of the zero-bias anomaly.}
  \end{figure}

We investigated mechanically-controllable break junctions at 1\,K in vacuum. 
Lower temperatures down to 0.1\,K or higher ones up to 4.2\,K did not affect the results.
Cu is a non-magnetic normal metal and Fe a band-ferromagnet with 
$T_{Curie} = 1043\,$K \cite{Ashcroft1976}.
The samples were made of wires with $0.1 - 0.5\,$mm diameter. 
The wire was glued onto a flexible bending plate and a groove was cut into the 
wire with a sharp knife, abrasive paper, or a thin corundum blade to define the 
break position.
After installation the radii of the unbroken contacts were about $5\,\mu$m.
They had residual resistance ratios of approximately 25, thus contacts down  to 
about $1\,\Omega$ should be in the ballistic limit.
Figure \ref{kondo-fit} shows the spectra of a) a low-resistance and b) a 
high-resistance Fe contact together with th  definition of the various parameters.

\begin{figure}
  \includegraphics{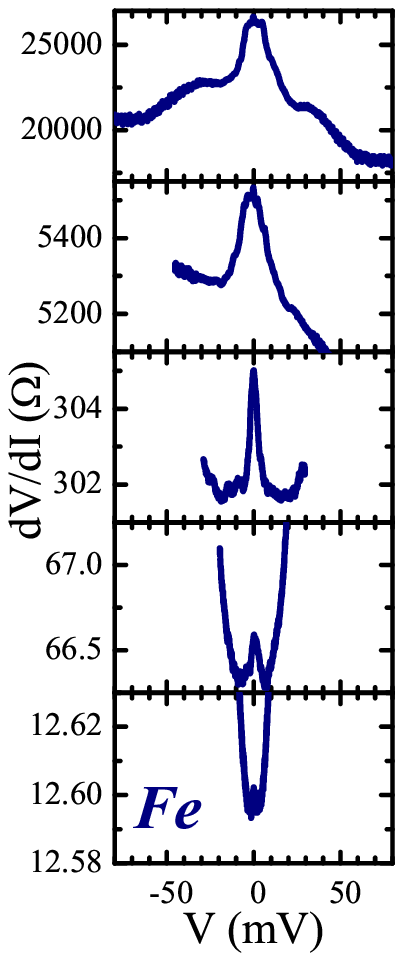}
  \includegraphics{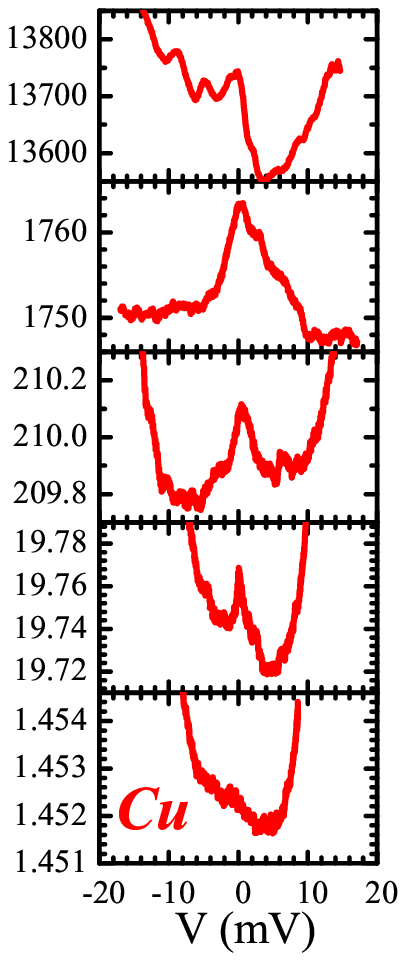}
  \caption{\label{spectra}
  Typical $dV/dI(V)$ spectra of Fe and Cu break junctions at 1\,K.
  To obtain the magnitude $\delta R$ and width $\delta V$ of the zero-bias anomaly
  the spectra were usually symmetrized first.}
  \end{figure}

Figure \ref{spectra} displays two series of typical spectra of Fe and Cu
contacts.
Low-resistance contacts usually showed the zero-bias anomaly, unless it 
was hidden in the background noise, together with the typical features of 
electron-phonon interaction that indicate ballistic transport. 
When the contact radius was reduced and the resistance increased, the
relative magnitude of the zero-bias anomaly grew and, at the same time, the 
spectroscopic features of electron-phonon interaction became suppressed.
In few cases we found an inverted zero-bias anomaly, a minimum, of 
comparable size as the maximum. 
Its origin is unclear - possibly it is caused by an accidental fabrication of a tunnel junction - 
and because it was only rarely observed we could not study it in detail.
Very often the zero bias peaks showed multiple sub-structures, like some of the 
contacts in Figure \ref{spectra}, that looked like resonances.

Figure \ref{magnitude} summarizes the data. 
Over a wide range of contact resistances the magnitude of the zero-bias anomaly
$\delta R \sim R^{2}$ for Fe
while Cu has a systematically weaker $\delta R \sim R^{3/2}$ dependence. 
These dependencies contrast the typical spectroscopic features like the 
electron-phonon interaction which vary as $\delta R \sim \sqrt{R}$ according 
to Equation \ref{pc-backscatter}.
There is no noticeable transition between small and large contacts,
indicating that the tiny zero-bias anomalies at small $R$ 
develop directly into the huge anomalies at large $R$. Thus the same 
mechanism is responsible for those anomalies, independent of the contact size.
Within the scattering of data points in Figure \ref{magnitude}, one can 
also recognize a trend towards larger widths when the resistance increases.

\section{Discussion}

\begin{figure}
  \includegraphics{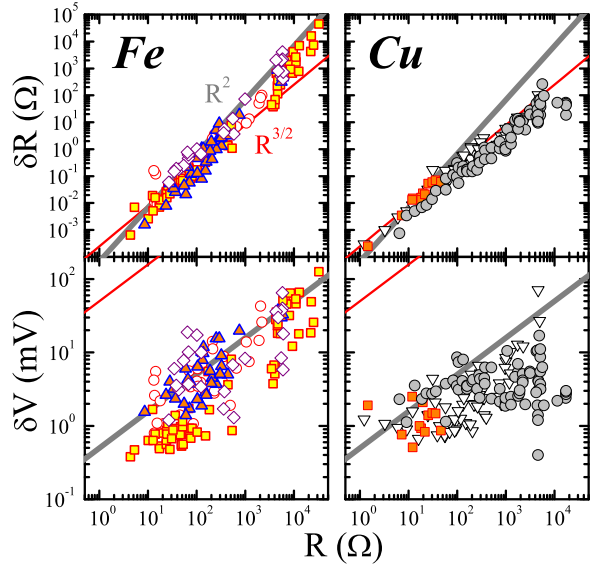}
  \caption{\label{magnitude}
    Magnitude $\delta R$ as well as width $\delta V$ of the zero-bias anomaly 
    as function of resistance $R$ of Fe and Cu break junctions at 1\,K.
    The thick solid lines are $\delta R = (9\pi / 16) R^2/R_K$ and
   $\delta V = 2\cdot \hbar v_F / e\xi_K$ when the Kondo length equals 
    $\xi_K = 100 \cdot a$ as discussed in the text. 
   The thin solid lines indicate a smaller slope of $R \sim R^{3/2}$
    and a Kondo length $\xi_K = a$.
    Different symbols indicate different series of measurements.}
  \end{figure}

A single impurity in the unitary limit contributes  $\delta R = 
\left(9\pi R^2 \right) / \left( 16 R_K \right)$
to the magnitude of the zero-bias anomaly \cite{Gloos2009} 
 - the same as switching of a single conducting channel.
This describes the experimental data for Fe in Figure  \ref{magnitude} quite well.
The width of the anomalies varies between 1\,mV for large contacts and
100\,mV for small contacts, implying Kondo lengths between
$1\,\mu$m and 10\,nm. 
This is about 100 times larger than the contact radius and supports the idea 
that a single impurity is responsible for the anomalies \cite{Gloos2009}.

Kondo phenomena in small-scale devices have been observed, for example, 
for tunneling into a single magnetic atom on a metallic surface \cite{Vitali2008}
or in quantum-point contacts and quantum dots in a two-dimensional electron 
gas, where a single electron sits either in the dot 
\cite{Meir1993,Goldhaber1998,Cronenwett1998,Bird2004} or in a shallow 
potential minimum near the center of the point contact \cite{Rejec2006}.
Transport through the constriction depends then on the spin of this single electron.
We have suggested a similar picture that one or few electrons are trapped near 
the contact and polarize the conduction electrons in the contact region \cite{Gloos2009}.
Another scenario has been suggested recently for atomic-size contacts 
of the ferromagnets Fe, Co, and Ni  that had only few conducting channels and 
resistances near $R_K$. There the zero-bias anomalies were attributed to the Kondo 
effect caused by the changed band structure at the contact \cite{Calvo2009}. 
Whether this could play a role in our experiments is unclear.

\section{Conclusions}
We have found a reproducible dependence of the magnitude of the 
zero-bias anomalies as function of contact size for Fe as well as for Cu.
If we accept that the Kondo effect is possible at atomic-size Fe contacts
because ferromagnetism is quenched by the small size, it should also
be possible in junctions of Cu which is {\it per se} non-magnetic. 
Making the contacts larger does not suppress the Kondo effect, as it survives
at least till the radii become larger than about 15\,nm.

\section{Acknowledgements}
We thank the Jenny and Antti Wihuri Foundation and the Magnus Ehrnroot 
Foundation for financial support.


\end{document}